Research article

# Interface coupling properties and reflection of bulk spin waves from biaxial multilayer ferromagnetic media

Sergey A Reshetnyak* and Tat'yana A Homenko[†]

Address: National Technical University of Ukraine, Kyiv Polytechnic Institute, Kyiv, 03057, Peremohy av. 37, Ukraine

Email: Sergey A Reshetnyak* - rsa@users.ntu-kpi.kiev.ua; Tat'yana A Homenko - homenko_t@ukr.net

* Corresponding author    †Equal contributors

## Abstract

The reflection coefficient of bulk spin waves from multilayer ferromagnetic structure with periodically modulated parameters of exchange interaction, uniaxial and rhombic magnetic anisotropy and saturation magnetization is calculated with a non-ideal coupling between layers. The strong dependence of spin-wave reflection coefficient on frequency, magnetic field and parameter of interfacial coupling is revealed. It allows changing the reflection intensity from 0 to 1 by changing either only external magnetic field value or frequency. The proposed model of boundary conditions gives the opportunity to take into account the quality of interfaces when studying the reflection processes of spin waves in multilayer structures.

## Background

The physics of multilayer structures is a rapidly developing branch of science that promises ample opportunities for applications of such objects, in particular, in microelectronics. Thereupon, characterizing the propagation of spin waves in such structures has generated heightened interest [1-3], having led to a new research area of magnetism – magnonics [4]. In published papers [5-7], the spectrum of bulk spin waves and reflection of bulk spin waves were investigated with the assumption of ideal exchange boundary conditions between layers of infinite multilayers with uniaxial magnetic structure. The present paper is devoted to studying the reflective characteristics of a multilayer biaxial ferromagnetic media taking into account non-ideal magnetic properties of interfaces between layers, which lead to a "defective" coupling interaction at interfaces. In addition to modulation of the bulk exchange interaction and anisotropy, modulation of the saturation magnetization will be considered.

## Methods

We use the equation of magnetic moment dynamics is used at the parameterization of spin density to describe the behaviour of spin waves in the considered system. Calculations are carried out at the continuous approach and in the exchange mode.

## Results

### *The basic equations*

Consider a system consisting of three parts, whose planes of contact are parallel to the *yz* plane. The first and third parts (in the direction of *x*-axis) are homogeneous, biaxial, semi-infinite ferromagnetic media, with an *N*-layer ferromagnetic structure between them. The modulated parameters are exchange interaction, $\alpha$, uniaxial, $\beta$, and rhombic, $\rho$, magnetic anisotropy, and saturation magnetization, $M_0$. As shown in figure 1, each layer of the *N*-layer ferromagnetic structure consists of two layers with thicknesses *a* and *b*. Parameters $\alpha$, $\beta$, $\rho$ and $M_0$ have values $\alpha_1$, $\beta_1$, $\rho_1$, $M_{01}$ and $\alpha_2$, $\beta_2$, $\rho_2$, $M_{02}$ in corresponding layers. The easy axis is parallel to an external homogeneous permanent magnetic field, $H_0$, itself parallel to *z*-axis.

Using the formalism of spin density [8], the magnetization can be written as

$$\mathbf{M}_j(\mathbf{r},t) = M_{0j} \Psi_j^+(\mathbf{r},t) \boldsymbol{\sigma} \Psi_j(\mathbf{r},t), \quad j = 1, 2, \qquad (1)$$

where $\Psi_j$'s are quasiclassical wave functions playing the role of a spin density order parameter, **r** is a radius-vector of the Cartesian coordinate system, *t* is time, and $\sigma$ are Pauli matrices.

Write Lagrange's equations for $\Psi_j$ as

$$i\hbar \frac{\partial \Psi_j(\mathbf{r},t)}{\partial t} = -\mu_0 \mathbf{H}_{ej}(\mathbf{r},t) \boldsymbol{\sigma} \Psi_j(\mathbf{r},t), \qquad (2)$$

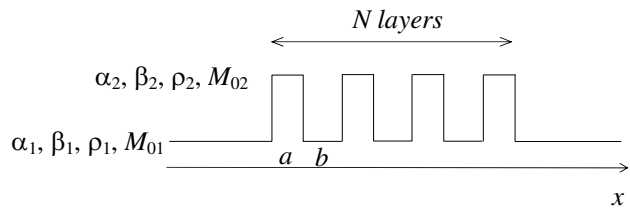

**Figure 1**
Change of magnetic parameters along the structure.

where $\mu_0$ is a Bohr magneton, $\mathbf{H}_{ej} = -\frac{\partial w_j}{\partial \mathbf{M}_j} + \frac{\partial}{\partial x_k}\frac{\partial w_j}{\partial(\partial \mathbf{M}_j/\partial x_k)}$, and $w_j$ is energy density. Note that, in an exchange mode when the condition $L \gg l = a + b$ is satisfied (here $L$ is material's characteristic length), energy density has the following form in each of homogeneous parts,

$$w_j = \frac{\alpha}{2}\left(\frac{\partial m_j}{\partial x_k}\right)^2 + \frac{\beta}{2}\left(m_{jx}^2 + m_{jy}^2\right) + \frac{\rho}{2}m_{jx}^2 - H_0 M_{jz}, \quad j = 1, 2. \tag{3}$$

Here it is taken into account, that the material is magnetized parallel to $\mathbf{e}_z$ in the ground state, $M_j^2(\mathbf{r}, t) = \text{const}$ and $\mathbf{M}_j(\mathbf{r}, t) = M_{0j}\mathbf{e}_z + \mathbf{m}_j(\mathbf{r}, t)$, where $\mathbf{m}_j(\mathbf{r}, t)$ is a small deviation of the magnetization from the ground state.

Then, using the linear perturbation theory, it is possible to write down the solution of (2) as

$$\Psi_j(\mathbf{r},t) = \exp(i\mu_0 H_0 t/\hbar)\cdot\begin{pmatrix} 1 \\ \chi_j(\mathbf{r},t) \end{pmatrix}, \tag{4}$$

where $\chi_j(\mathbf{r}, t)$ is a small function characterizing the deviation of a magnetization from the ground state. Linearizing equation (2) while taking into account equation (4), we obtain the following equation,

$$\begin{aligned} -\frac{i\hbar}{2\mu_0 M_{0j}}\frac{\partial \chi_j(\mathbf{r},t)}{\partial t} &= \left(\alpha_j\Delta - \beta_j - \frac{\rho_j}{2} - \tilde{H}_{0j}\right)\chi_j(\mathbf{r},t) - \frac{\rho_j}{2}\chi_j^*(\mathbf{r},t), \\ \frac{i\hbar}{2\mu_0 M_{0j}}\frac{\partial \chi_j^*(\mathbf{r},t)}{\partial t} &= \left(\alpha_j\Delta - \beta_j - \frac{\rho_j}{2} - \tilde{H}_{0j}\right)\chi_j^*(\mathbf{r},t) - \frac{\rho_j}{2}\chi_j(\mathbf{r},t), \end{aligned} \tag{5}$$

where $\tilde{H}_{0j} = H_0/M_{0j}$.

Expressing $\chi^*(\mathbf{r}, t)$ from the first of these equations and substituting it into the second, we obtain the following equation for magnetization dynamics:

$$\begin{aligned} -\frac{\hbar^2}{(2\mu_0 M_{0j})^2}\frac{\partial^2 \chi_j(\mathbf{r},t)}{\partial t^2} &= \\ &= \left[\alpha_j^2\Delta^2 - 2\alpha_j\left(\beta_j + \frac{\rho_j}{2} + \tilde{H}_{0j}\right)\Delta + (\beta_j + \tilde{H}_{0j})(\beta_j + \rho_j + \tilde{H}_{0j})\right]\chi_j(\mathbf{r},t). \end{aligned} \tag{6}$$

Performing the Fourier transformations on $y$ and $z$ coordinates and time, $t$, we can write

$$\left[\alpha_j^2 \frac{\partial^4}{\partial x^4} - 2\alpha_j \left(\beta_j + \frac{\rho_j}{2} + \tilde{H}_{0j}\right)\frac{\partial^2}{\partial x^2} + \alpha_j^2 k_{j\perp}^4 + 2\alpha_j k_{j\perp}^2 \left(\beta_j + \frac{\rho_j}{2} + \tilde{H}_{0j}\right) + \right.$$
$$\left. + \left(\beta_j + \tilde{H}_{0j}\right)\left(\beta_j + \rho_j + \tilde{H}_{0j}\right) - \Omega_j^2\right]\chi_j^{\omega,\mathbf{k}_\perp}(x) = 0. \quad (7)$$

Here, $\Omega_j = \omega\hbar/2\mu_0 M_{0j}$, $\omega$ is frequency, and $\mathbf{k}_\perp = (0, k_y, k_z)$.

The spin wave reflection amplitude from an $N$-layer structure can be represented as [9],

$$R_N = R\frac{1-\exp(2iqlN)}{1-R^2\exp(2iqlN)}, \quad (8)$$

where $R$ is an amplitude of reflection from semi-infinite multilayer structure ($N = \infty$),

$$R = \frac{\sqrt{(r+1)^2-\tau^2}-\sqrt{(r-1)^2-\tau^2}}{\sqrt{(r+1)^2-\tau^2}+\sqrt{(r-1)^2-\tau^2}}, \quad (9)$$

$q$ is a Bloch wave vector defined by,

$$\exp(iqlN) = \frac{\sqrt{(\tau+1)^2-r^2}+\sqrt{(\tau-1)^2-r^2}}{\sqrt{(\tau+1)^2-r^2}-\sqrt{(\tau-1)^2-r^2}}, \quad (10)$$

$l = a + b$ is the structure period, $r$ and $\tau$ are the complex amplitudes of reflection and transmission accordingly for a single symmetric (with regard to its center) period.

Since the equations (5) have the form similar to Schrödinger equation, amplitudes of reflection and transmission for a single period can be found using the corresponding method of quantum mechanics.

### *Boundary conditions*

For a material consisting of two homogeneous parts in contact along the $yz$ plane, it is possible to write down the energy density as,

$$w = \sum_{j=1}^{2}\theta\left[(-1)^j x\right]w_j + A\delta(x)\mathbf{M}_1\mathbf{M}_2, \quad (11)$$

where $A$ is the constant describing a coupling along the interface, $\theta(x)$ is a step function, and $w_j$'s are defined by (3). After integrating the equations of movement of magnetic moment in the vicinity of an interface, we obtain the following boundary conditions (indexes $\omega$, $\mathbf{k}_\perp$ are omitted),

$$A\gamma(\chi_2 - \chi_1) + \alpha_1 \chi'_1 = 0,$$
$$A(\chi_2 - \chi_1) + \gamma\alpha_2 \chi'_2 = 0, \quad (12)$$

where $\gamma = M_{02}/M_{01}$, and prime means the derivative with $x$. These boundary conditions will be applied at each interface of the multilayer structure.

### *Amplitudes of reflection and transmission for a single period*

Represent an incident wave with function $\chi_I = \exp(ik_1 x)$, a reflected wave with function $\chi_r = r \exp(i\tilde{k}_1 x)$, and a wave transmitted through a single layer with function $\chi_\tau = \tau \exp(ik_1 x)$. Here $k_1$ and $\tilde{k}_1$ are the wave vectors of incident and reflected waves and accordingly $\tilde{k}_1 = -k_1$. Substituting these expressions into (12) together with the expression $\chi_{layer} = C_1 \exp(ik_2 x) + C_2 \exp(-ik_2 x)$ describing the wave within the intermediate layer, for each of two borders of a single period we come to the expressions for spin wave amplitudes of reflection and transmission,

$$r = \exp(ik_1 b) \cdot \frac{A^2 C_- + AD_- + E}{A^2 C_+ + AD_+ + E},$$
$$\tau = \exp(-ik_1 a) \cdot \frac{2A^2 \alpha_1 \alpha_2 k_1 k_2}{A^2 C_+ + AD_+ + E}, \quad (13)$$

where

$$C_\pm = (1 \pm \gamma^4)\alpha_1 \alpha_2 k_1 k_2 \cos(k_2 a) - i\gamma^2 (\alpha_1^2 k_1^2 \pm \alpha_2^2 k_2^2) \sin(k_2 a),$$
$$D_\pm = \gamma\alpha_1 \alpha_2 k_1 k_2 \left[ (1 \pm \gamma^2)\alpha_2 k_2 \sin(k_2 a) + i(1 + \gamma^2)\alpha_1 k_1 \cos(k_2 a) \right],$$
$$E = i\gamma^2 \alpha_1^2 \alpha_2^2 k_1^2 k_2^2 \sin(k_2 a),$$
$$\alpha_j k_j^2 = \sqrt{\Omega_j^2 + \rho_j^2/4} - \beta_j - \rho_j/2 - \alpha_j k_{j\perp}^2 - \tilde{H}_{0j}.$$

### *Amplitudes of reflection and transmission for multilayer structure*

Using expression (13), it is possible to rewrite the equation (9) as

$$R = \frac{\sqrt{F_+ F_-} - \sqrt{G_+ G_-}}{\sqrt{F_+ F_-} + \sqrt{G_+ G_-}}, \quad (14)$$

where

$$F_{\pm} = A^2 \left\{ \left[ \cos(k_1 b/2) - i\gamma^4 \sin(k_1 b/2) \pm 1 \right] \cos(k_2 a) - \right.$$
$$- \left[ \left(\gamma^2 \zeta \pm 1\right) \sin(k_1 b/2) + i\left(\gamma^2 \zeta^{-1} \pm 1\right) \cos(k_1 b/2) \right] \sin(k_2 a) \right\} +$$
$$+ A\gamma\alpha_1 k_1 \left\{ \zeta \left[ \cos(k_1 b/2) - i\gamma^2 \sin(k_1 b/2) \right] \sin(k_2 a) + i\left(1 + \gamma^2\right) \cos(k_1 b/2) \cos(k_2 a) \right\} +$$
$$+ i\gamma^2 \alpha_1^2 k_1^2 \zeta \cos(k_1 b/2) \sin(k_2 a),$$

$$G_{\pm} = A^2 \left\{ \left[ i\sin(k_1 b/2) - \gamma^4 \cos(k_1 b/2) \mp 1 \right] \cos(k_2 a) + \right.$$
$$+ \left[ \left(\gamma^2 \zeta^{-1} \pm 1\right) \sin(k_1 b/2) + i\left(\gamma^2 \zeta \pm 1\right) \cos(k_1 b/2) \right] \sin(k_2 a) \right\} +$$
$$+ A\gamma\alpha_1 k_1 \left\{ \zeta \left[ i\sin(k_1 b/2) - \gamma^2 \cos(k_1 b/2) \right] \sin(k_2 a) - \left(1 + \gamma^2\right) \sin(k_1 b/2) \cos(k_2 a) \right\} -$$
$$- \gamma^2 \alpha_1^2 k_1^2 \zeta \sin(k_1 b/2) \sin(k_2 a),$$

$$\zeta = \alpha_2 k_2 / \alpha_1 k_1.$$

The reflection amplitude for a multilayer structure consisting of *N* layers, is defined by expression (8).

Note that as calculating the reflection amplitude (14) requires extracting the square root of complex expressions, there will be branching points, whenever the following conditions are met,

$$F_+ F_- = 0,$$

$$G_+ G_- = 0.$$

## Discussion

Figures 2, 3 and 4 illustrate dependence of the reflection coefficients $|R|^2$ and $|r|^2$ on frequency at fixed value of the magnetic field for three values of *A* and parameters characteristic of ferrogarnets [10]. The intensity of the reflected wave depends on frequency. Both the forbidden zones, and the periodically repeating points corresponding to full transmission of a wave through the multilayer structure.

Figures 5, 6, 7 and 8 show the field dependence at a fixed frequency. As can seen on these diagrams the process of displacement of characteristic peaks to smaller magnetic field values with reduction of parameter A. Besides, similar dependences allow changing values of reflection coefficients in a wide range by means of change of value of an external magnetic field whereas the material's parameters and frequency are fixed.

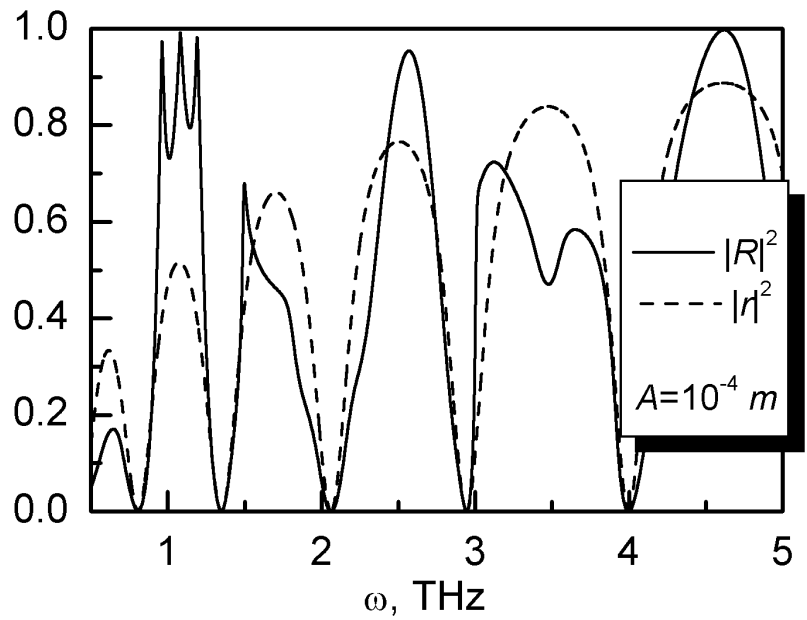

**Figure 2**
**Dependence of the reflection coefficients $|R|^2$ and $|r|^2$ on frequency, $A = 10^{-4}$ m**. $\alpha_1 = 10^{-11}$ $m^2$, $\alpha_2 = 2\cdot 10^{-11}$ $m^2$, $\beta_1 = 40$, $\beta_2 = 90$, $\rho_1 = 5$, $\rho_2 = 10$, $M_{01} = 90$ G, $M_{02} = 95$ G, $H_0 = 2.3$ $\kappa Oe$, $A = 10^{-4}$m, $a = 2\cdot 10^{-6}$m, $b = 10^{-6}$m.

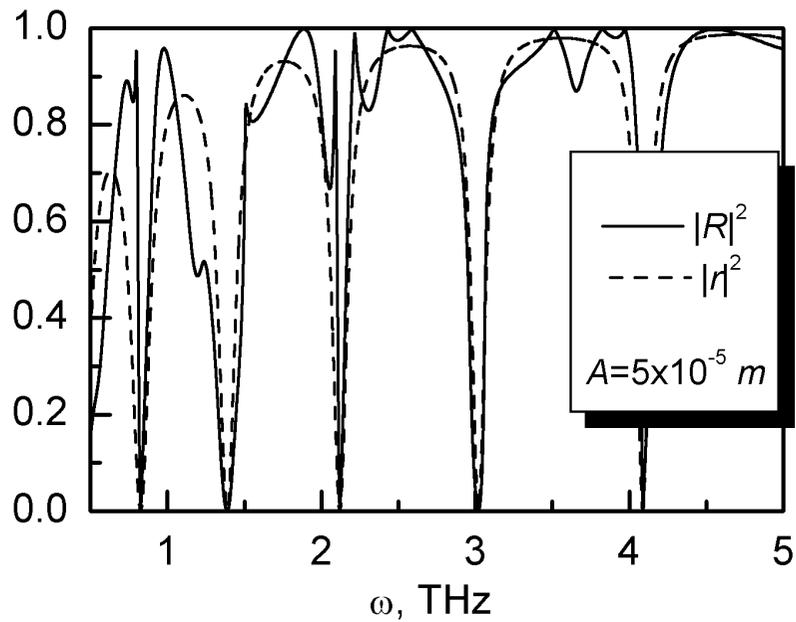

**Figure 3**
**Dependence of the reflection coefficients $|R|^2$ and $|r|^2$ on frequency, $A = 5\cdot 10^{-5}$m**. $\alpha_1 = 10^{-11}$ $m^2$, $\alpha_2 = 2\cdot 10^{-11}$ $m^2$, $\beta_1 = 40$, $\beta_2 = 90$, $\rho_1 = 5$, $\rho_2 = 10$, $M_{01} = 90$ G, $M_{02} = 95$ G, $H_0 = 2.3$ $\kappa Oe$, $A = 10^{-5}$m, $a = 2\cdot 10^{-6}$m, $b = 10^{-6}$m.

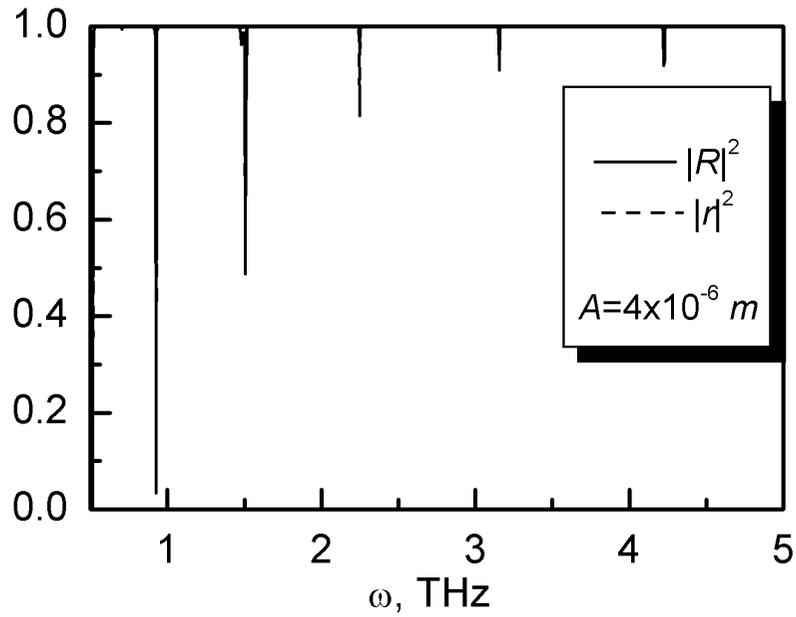

**Figure 4**
**Dependence of the reflection coefficients $|R|^2$ and $|r|^2$ on frequency, $A = 4\cdot10^{-6}m$**. $\alpha_1 = 10^{-11}\ m^2$, $\alpha_2 = 2\cdot10^{-11}\ m^2$, $\beta_1 = 40$, $\beta_2 = 90$, $\rho_1 = 5$, $\rho_2 = 10$, $M_{01} = 90\ G$, $M_{02} = 95\ G$, $H_0 = 2.3\ \kappa Oe$, $A = 10^{-6}m$, $a = 2\cdot10^{-6}m$, $b = 10^{-6}m$.

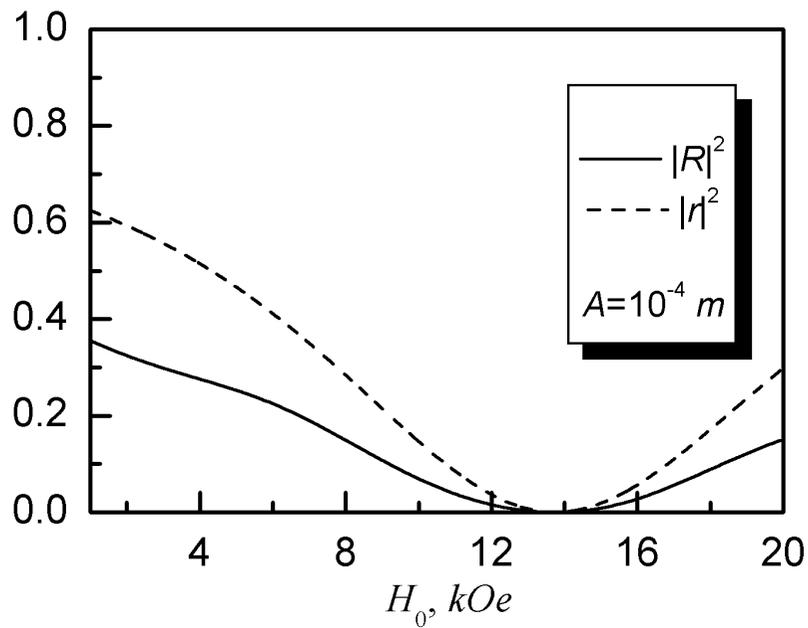

**Figure 5**
**Dependence of the reflection coefficients $|R|^2$ and $|r|^2$ on external permanent magnetic field, $A = 10^{-4}m$**. $\alpha_1 = 10^{-11}\ m^2$, $\alpha_2 = 2\cdot10^{-11}\ m^2$, $\beta_1 = 40$, $\beta_2 = 90$, $\rho_1 = 5$, $\rho_2 = 10$, $M_{01} = 90\ G$, $M_{02} = 95\ G$, $\omega = 2.3\ THz$, $A = 10^{-4}m$, $a = 2\cdot10^{-6}m$, $b = 10^{-6}m$.

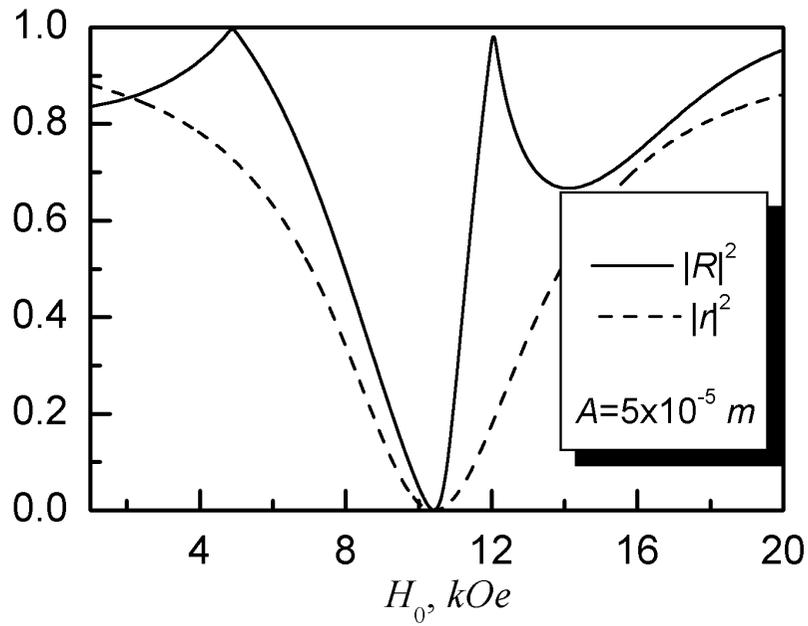

**Figure 6**
**Dependence of the reflection coefficients |R|² and |r|² on external permanent magnetic field, A = 5·10⁻⁵ m**. $\alpha_1 = 10^{-11}\,m^2$, $\alpha_2 = 2\cdot 10^{-11}\,m^2$, $\beta_1 = 40$, $\beta_2 = 90$, $\rho_1 = 5$, $\rho_2 = 10$, $M_{01} = 90\,G$, $M_{02} = 95\,G$, $\omega = 2.3\,THz$, $A = 10^{-5}m$, $a = 2\cdot 10^{-6}m$, $b = 10^{-6}m$.

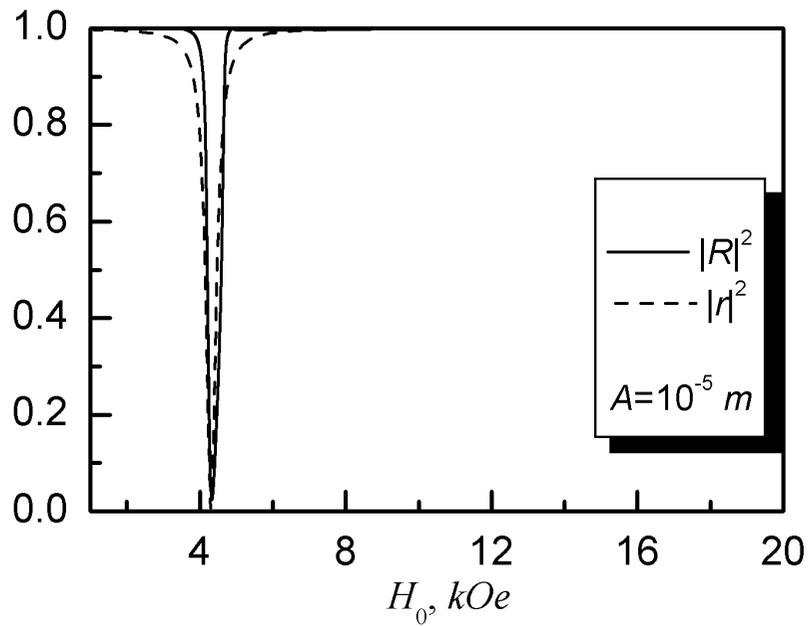

**Figure 7**
**Dependence of the reflection coefficients |R|² and |r|² on external permanent magnetic field, A = 10⁻⁵ m**. $\alpha_1 = 10^{-11}\,m^2$, $\alpha_2 = 2\cdot 10^{-11}\,m^2$, $\beta_1 = 40$, $\beta_2 = 90$, $\rho_1 = 5$, $\rho_2 = 10$, $M_{01} = 90\,G$, $M_{02} = 95\,G$, $\omega = 2.3\,THz$, $A = 10^{-5}m$, $a = 2\cdot 10^{-6}m$, $b = 10^{-6}m$.

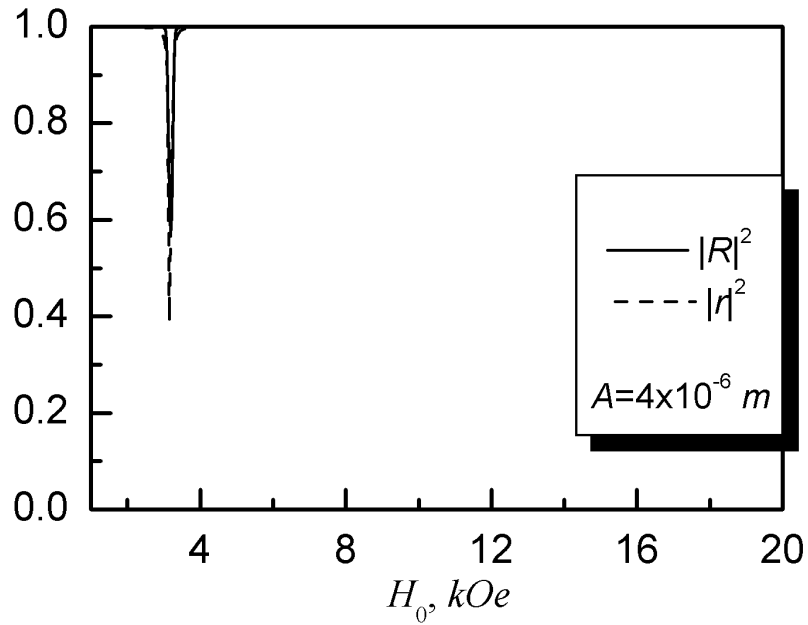

**Figure 8**
**Dependence of the reflection coefficients $|R|^2$ and $|r|^2$ on external permanent magnetic field, $A = 4 \cdot 10^{-6}$m**. $\alpha_1 = 10^{-11} m^2$, $\alpha_2 = 2 \cdot 10^{-11} m^2$, $\beta_1 = 40$, $\beta_2 = 90$, $\rho_1 = 5$, $\rho_2 = 10$, $M_{01} = 90\ G$, $M_{02} = 95\ G$, $\omega = 2.3\ THz$, $A = 10^{-6}m$, $a = 2 \cdot 10^{-6}m$, $b = 10^{-6}m$.

Figure 9 shows dependencies of reflection coefficients $|R|2$ and $|r|2$ on parameter A at fixed frequency and fixed external magnetic field. Note that the case A = 0 is equivalent to the absence of an exchange of layers through an interface (this fact leads to full reflection $|R|2 = |r|2 = 1$ for all frequencies including points of resonance), and $A \to \infty$ corresponds to an ideal (in a coupling sense) boundary. Thus, when $A \to \infty$, we obtain usual coupling boundary conditions [11]. In general, a change in the value of parameter A can be interpreted as change of effective distance between adjacent layers due to what an interlayer exchange or increases ($A \to \infty$) or decreases ($A \to 0$). Parameter A can be estimated as being on the order of $A \sim \alpha/d$, where d is effective thickness of an interface.

The dependencies obtained for spin wave reflection from a three-layer structure ($N = 3$) practically coincide with the dependencies corresponding to amplitude $R$ of reflection from a semi-infinite multilayer structure for the adduced material's parameters, therefore only the diagrams corresponding to $|R|^2$ and $|r|^2 \equiv |R_1|^2$ are shown in the figures.

## Conclusion

Thus, the reflective ability of multilayer structure not only has strong dependence on frequency and external field, but also is defined in many respects by value of parameter *A* on boundaries, and it is manifested especially brightly at small values of *A*. It is necessary to take into account this fact and make the corresponding amendment on an irregularity of an exchange in the interface at development of filters and other devices of spin-wave microelectronics.

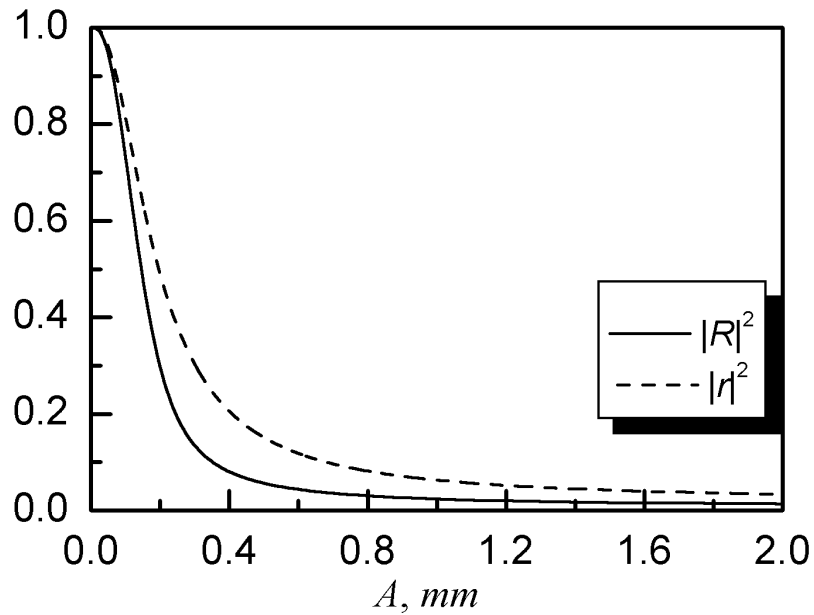

**Figure 9**
**Dependence of the reflection coefficients |R|² and |r|² on interfacial coupling parameter**. $\alpha_1 = 10^{-11}\ m^2$, $\alpha_2 = 2\cdot 10^{-11}\ m^2$, $\beta_1 = 40$, $\beta_2 = 90$, $\rho_1 = 5$, $\rho_2 = 10$, $M_{01} = 90\ G$, $M_{02} = 95\ G$, $\omega = 4.3\ THz$, $H_0 = 2.3\ \kappa Oe$, $a = 2\cdot 10^{-6} m$, $b = 10^{-6} m$.

## References

1. Skubic B, Holmström E, Eriksson O: *Phys Rev B* 2004, **70:**094421. 9 pages
2. Zhang J, Levy PM: *Phys Rev B* 2004, **70:**184442. 6 pages
3. Gu Y, Zhang D, Zhan X, Ji Z, Zhang Y: *J Magn Magn Mater* 2006, **297:**7-16.
4. Kruglyak VV, Hicken RJ: *J Magn Magn Mater* 2006, **30:**191-194.
5. Gorobets YI, Zubanov AE, Kuchko AN, Shedzhuri KD: *Physics of the Solid State* 1992, **34:**790-792. *Fiz Tverd Tela* 1992, 34: 1486–1490
6. Kruglyak VV, Kuchko AN: *Physica B* 2003, **339:**130-133.
7. Gorobets YI, Kuchko AN, Reshetnyak SA: *Physics of the Solid State* 1996, **38:**315-317. *Fiz Tverd Tela* 1992, 38: 575–580
8. Bar'yakhtar VG, Gorobets YI: *Bubble domains and their lattices* Kyiv: Naukova dumka; 1988.
9. Ignatovich VK: *Sov Phys Uspekhi* 1986, **150:**880-887. Uspekhi Fiz Nauk 1986, 150: 145–158
10. Eshenfelder A: *Magnetic bubble technology* New York: Springer-Verlag; 1980.
11. Akhiezer AI, Bar'yakhtar VG, Peletminski SV: *Spin waves* Moscow: Nauka; 1967.